\documentclass[prb,twocolumn,aps,showpacs,fixfloats]{revtex4}
\usepackage{graphicx}
\usepackage{bm}
\usepackage{amsmath,amssymb}
\usepackage{subfigure}
\usepackage{float}
\usepackage{latexsym}
\usepackage{epstopdf}
\usepackage{color}
\usepackage{enumerate}
\usepackage{pdfpages}

\begin{document}
\newcommand{\s}{\scriptscriptstyle}
\newcommand{\uu}{\uparrow \uparrow}
\newcommand{\ud}{\uparrow \downarrow}
\newcommand{\du}{\downarrow \uparrow}
\newcommand{\dd}{\downarrow \downarrow}
\newcommand{\ket}[1] { \left|{#1}\right> }
\newcommand{\bra}[1] { \left<{#1}\right| }
\newcommand{\bracket}[2] {\left< \left. {#1} \right| {#2} \right>}
\newcommand{\vc}[1] {\ensuremath {\bm {#1}}}
\newcommand{\tr}{\text{Tr}}
\newcommand{\Trans}{\ensuremath \Upsilon}
\newcommand{\Refl}{\ensuremath \mathcal{R}}
\renewcommand{\u}{\uparrow}
\renewcommand{\d}{\downarrow}

\newcommand{\Pcyl}{\mathcal{P}_{\s \textit{cyl}}}
\newcommand{\Psph}{\mathcal{P}_{\s \textit{sphere}}}


\title {Spin injection from a ferromagnet into a semiconductor in the case of a rough interface}

\author{R. C. Roundy  and M. E. Raikh}

 \affiliation{ Department of Physics and
Astronomy, University of Utah, Salt Lake City, UT 84112, USA}

\begin{abstract}
The effect of the interface roughness on the spin injection from a ferromagnet into a semiconductor is studied theoretically. Even a  small interface irregularity can lead to a significant enhancement of the injection efficiency.
When a typical size of the irregularity, $a$, is within a domain $\lambda_{\s F} \ll a \ll \lambda_{\s N}$, where
$\lambda_{\s F}$ and $\lambda_{\s N}$ are the spin-diffusion lengths in the ferromagnet and semiconductor, respectively,
the geometrical enhancement factor is $\sim \lambda_{\s N}/a$.  The origin of the enhancement is the modification of the local electric field on small scales $\sim a$ near the interface. We demonstrate the effect of enhancement by considering
a number of analytically solvable examples of injection through curved ferromagnet-semiconductor interfaces. For a generic curved interface the
enhancement factor is  $\sim \lambda_{\s N}/R$, where $R$ is the local radius of curvature.
\end{abstract}


\maketitle

\section{Introduction}
\begin{figure}
\includegraphics[width=77mm]{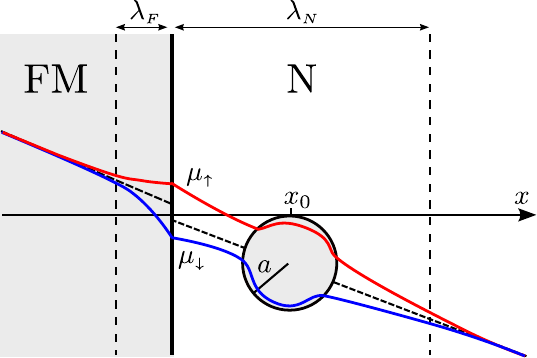}
\caption{(Color online) Illustration of the spin injection through a planar interface, Ref. \onlinecite{wyder}, in the presence of a ferromagnetic cylinder with a small radius $a\ll \lambda_{\s N}$ located at distance, $x_0$, such that $a\ll x_0 \ll \lambda_{\s N}$ from the F/N interface. The behaviors of the chemical potentials for up and down spins is shown schematically.    }
\label{fig1}
\end{figure}

In a seminal paper Ref. \onlinecite{wyder} the efficiency of spin injection
from a ferromagnet into a normal metal was quantified.
In the minimal model considered in Ref. \onlinecite{wyder} spin injection resulted
from the behavior of four functions $\mu^{\s F}_{\s \u}(x), \mu^{\s F}_{\s \d}(x), \mu^{\s N}_{\s \u}(x)$ and $\mu^{\s N}_{\s \d}(x)$,
which are the chemical potentials for the spin-up and spin-down electrons near the boundary.
In the ferromagnetic region, $x<0$, and in the metallic region, $x>0$, see Fig. \ref{fig1}, the functions $\mu_{\s \u}$ and  $\mu_{\s \u}$
are related by the diffusion equations
\begin{equation}
\frac{d^2}{dx^2} \left( \mu^{\s F}_{\s \u} - \mu^{\s F}_{\s \d} \right)
= \frac{\mu^{\s F}_{\s \u} - \mu^{\s F}_{\s \d}}{\lambda_{\s F}^2},
\frac{d^2}{dx^2} \left( \mu^{\s N}_{\s \u} - \mu^{\s N}_{\s \d} \right)
= \frac{\mu^{\s N}_{\s \u} - \mu^{\s N}_{\s \d}}{\lambda_{\s N}^2},
\label{diffusion-eq}
\end{equation}
where $\lambda_{\s F}$ and $\lambda_{\s N}$ are the spin diffusion lengths in the
ferromagnet and  normal metal, respectively.  The system Eq. (\ref{diffusion-eq}) should
be supplemented by the boundary conditions
\begin{align}
\left.  {j}^{\s F}_{\s \u} = \sigma^{\s F}_{\s \u} \frac{\partial \mu^{\s F}_{\s \u}}{\partial x} \right|_{x=0^{-}} &=
\left.  j^{\s N}_{\s \u} = \sigma^{\s N}         \frac{\partial \mu^{\s N}_{\s \u}}{\partial x} \right|_{x=0^{+}}, \nonumber \\
\left.  {j}^{\s F}_{\s \d} = \sigma^{\s F}_{\s \d} \frac{\partial \mu^{\s F}_{\s \d}}{\partial x} \right|_{x=0^{-}} &=
\left.  {j}^{\s N}_{\s \d} = \sigma^{\s N}         \frac{\partial \mu^{\s N}_{\s \d}}{\partial x} \right|_{x=0^{+}},
\label{current-condition}
\end{align}
which expresses the continuity of the  components of the current densities, $j_{\s \u}$ and $j_{\s \d}$, normal to the interface.  Here $\sigma^{\s F}_{\s \u}$ and $\sigma^{\s F}_{\s \d}$ are the
conductivities of up-spins and down-spins in the ferromagnetic region,
while $\sigma^{\s N}$ is the (spin independent) conductivity in the metallic region.
Another boundary condition, which follows from the continuity of $\mu_{\s \u}$ and $\mu_{\s \d}$
at the boundary $x=0$ reads
\begin{equation}
\left. \mu^{\s F}_{\s \u} - \mu^{\s F}_{\s \d} \right|_{x =0^{-}}
 = \left. \mu^{\s N}_{\s \u} - \mu^{\s N}_{\s \d} \right|_{x =0^{+}}.
\label{mu-condition}
\end{equation}
General solutions of Eq. (\ref{diffusion-eq}), which decay away from the boundary, are
the simple exponents
\begin{align}
\mu^{\s F}_{\s \u} &= \mu_0^{\s F}(x) + A_{\s \u} e^{x/\lambda_{\s F}},
 & \mu_{\s \u}^{\s N} &= \mu_0^{\s N}(x) + B_{\s \u} e^{-x/\lambda_{\s N}},  \\
\mu_{\s \d}^{\s F} &= \mu_0^{\s F}(x) + A_{\s \d} e^{x/\lambda_{\s F}},
& \mu_{\s \d}^{\s N} &= \mu_0^{\s N}(x) + B_{\s \d} e^{-x/\lambda_{\s N}},
\end{align}
where $\mu_0^{\s F}(x)$ and $\mu_0^{\s N}(x)$ are the linear functions with slopes proportional
to the net current, $j$; the ratio of the slopes is $(\sigma^{\s F}_{\s \u}+\sigma^{\s F}_{\s \d})/2\sigma^{\s N}$.
The conditions Eqs. (\ref{current-condition}), (\ref{mu-condition}) impose the following
relations between the coefficients $A_{\s \u}, A_{\s \d}, B_{\s \u}$, and $B_{\s \d}$

\begin{align}
\sigma^{\s F}_{\s \u} \left(  \frac{j}{\sigma^{\s F}_{\s \u} + \sigma^{\s F}_{\s \d}} + \frac{A_{\s \u}}{\lambda_{\s F}} \right)
&=\sigma^{\s N} \left( \frac{j}{2 \sigma^{\s N}}  - \frac{B_{\s \u}}{\lambda_{\s N}} \right) = j^{\s N}_{\s \u}, \\
\sigma^{\s F}_{\s \d} \left(  \frac{j}{\sigma^{\s F}_{\s \u} + \sigma^{\s F}_{\s \d}} + \frac{A_{\s \d}}{\lambda_{\s F}} \right)
&=\sigma^{\s N} \left( \frac{j}{2 \sigma^{\s N}}  - \frac{B_{\s \d}}{\lambda_{\s N}} \right) = j^{\s N}_{\s \d}, \\
A_{\s \u} - A_{\s \d} &= B_{\s \u} - B_{\s \d}.
\label{eq-cur}
\end{align}
These relations together with total current conservation, $j^{\s N}_{\s \u} + j^{\s N}_{\s \d} = j$, are sufficient to find the degree of  polarization of the injected current\cite{wyder}
\begin{equation}
\mathcal{P} = \frac{j^{\s N}_{\s \u} - j^{\s N}_{\s \d}}{j^{\s N}_{\s \u} + j^{\s N}_{\s \d}}
 =
\dfrac{ \sigma^{\s N}
\left( \dfrac{1}{\sigma^{\s F}_{\s \d}} - \dfrac{1}{\sigma^{\s F}_{\s \u}}\right) \gamma}{
2+ \sigma^{\s N}
\left( \dfrac{1}{\sigma^{\s F}_{\s \d}} + \dfrac{1}{\sigma^{\s F}_{\s \u}}\right) \gamma
},
\label{eq-wyder}
\end{equation}
where the parameter $\gamma$ is defined as
\begin{equation}
\label{gamma}
\gamma = \frac{\lambda_{\s F}}{\lambda_{\s N}}.
\end{equation}
It follows from Eq. (\ref{eq-wyder}) that the injection is efficient when the
conductivities of the ferromagnet and the normal metal are of the same order.
And indeed, in subsequent experiments\cite{vanWeesCointoAlNonlocal,MesoscopicIsland,CrossedGeometry1DCalculation}
the injection was demonstrated
for the contacts of ferromagnets with paramagnetic metals.

By the year 2000 it became apparent that applications of the spin-injection effect
in the information technology require the injection from a ferromagnet into a semiconductor.
Thus the subsequent experimental studies, see e.g. Refs. \onlinecite {tip, InjectionIntoNanowire, vanWeesGrapheneInjection, LatestFertNonlocal},
were focused on achieving this goal. The detailed
account of the results on  spin-injection devices based on ferromagnet-silicon contacts
can be found in a recent
review Ref. \onlinecite{reviewSi}.

It was first pointed out in Ref. \onlinecite{molenkamp} that the large
ratio of conductivities $\sigma^{\s F}/\sigma^{\s N} \sim 10^4$ constitutes
a fundamental obstacle for the spin injection limiting it to $\sim 0.1$
percent\cite{molenkamp}. To circumvent  this ``conductivity mismatch"
problem it was proposed\cite{rashba,review} to introduce a tunnel barrier
between the ferromagnet and semiconductor.  However Eq. (\ref{eq-wyder})
also suggests that the additional source of weakness of injection is the
the smallness of the parameter $\gamma$, Eq. (\ref{gamma}). While the
spin diffusion length in the ferromagnet is typically $\lambda_{\s F} \sim
10$ nm, the length $\lambda_{\s N}$ is much larger. From the  experiment on
injection into InN nanowires\cite{InjectionIntoNanowire} the value
$\lambda_{\s N}\sim 200$ nm was inferred.  Even higher values of $\lambda_{\s
N}$ between $5$ and $50$ $\mu$m have been reported for spin injection into
GaAs wires\cite{Crowell1,Crowell2}.

The main message of the present paper is that, with small value of $\gamma$, the polarization of the injected current
can be strongly enhanced by the roughness of the ferromagnet-semiconductor interface with a spatial scale $\sim \lambda_{\s F}$. On the qualitative level, this enhancement is due to the local enhancement of electric field near a
curved surface. Note that this effect could not be uncovered in the earlier theories of spin transport\cite{FertMultilayer, GorkovDzeroMultilayer, khaetskii, 5layers,Flatte, German} in multilayered structures, where it was implicit that the chemical potentials change along one dimension {\em only}.

In  Sections II and III we will illustrate our message for some toy models which allow a rigorous analytical solution.
In Sect. IV  we will consider the enhancement of injection due to interface inhomogeneities  for more realistic geometries. Sect. V concludes the paper.

\section{Ferromagnetic grain near the interface}

Assume that a ferromagnetic cylinder of radius, $a$, is embedded into a semiconducting region
at distance, $x_0$, from the interface. The distance $x_0$ is much bigger than $a$ but much smaller
than the spin diffusion length $\lambda_{\s N}$, see Fig. \ref{fig1}. The presence of the cylinder modifies
the current distribution in semiconductor. In principle, this modification depends on $\sigma_{\s \u}^{\s F}$ and $\sigma_{\s \d}^{\s F}$, but when both conductivities are bigger than $\sigma^{\s N}$, the correction to the current density
depends only on the distance, ${\bm \rho}$, to the center of the cylinder
\begin{equation}
\label{electric}
{\bm j}^{\s N}({\bm \rho}) = {\bm j}^{\s 0}  - {\bm j}^{\s 0} \frac{a^2}{\rho^2}
+ \frac{2 a^2 ({\bm j}^{\s 0} \cdot {\bm \rho}) {\bm \rho}}{\rho^4}.
\end{equation}
This textbook result emerges from matching the tangent components of electric field and normal components of current
at the surface of the cylinder. Our goal is to derive the relation similar to Eq. (\ref{electric}) for the {\em spin} current density
\begin{equation}
{\bm j}_{s}={\bm j}_{\s \u}-{\bm j}_{\s \d}.
\end{equation}
In the absence of the cylinder,
this density is given by ${\bm j}_{s}=\mathcal{P} {\bm j}^{\s 0} \exp(-x/\lambda_{\s N})$,
with $\mathcal{P}$ defined by Eq. (\ref{eq-wyder}).
To achieve this goal, one has to find the functions $\mu_{\s \u}^{\s N}$ and $\mu_{\s \d}^{\s N}$ from the diffusion
equation Eq. (\ref{diffusion-eq}) and match them and  the normal components of current with the corresponding solutions of the diffusion equation for $\mu_{\s \u}^{\s F}$ and $\mu_{\s \d}^{\s F}$ inside the cylinder.

For distances, $x_0$,  in the domain $a \ll x_0 \ll \lambda_{\s N}$  the solutions for
$\mu^{\s N}_{\s \u}$ and $\mu^{\s N}_{\s \d}$ still have the ``dipole'' form
\begin{align}
\mu_{\s \u}^{\s N} &= \alpha_{\s \u} + \beta_{\s \u} \rho \cos \theta
+ \frac{F}{\rho} \cos \theta, \\
\mu_{\s \d}^{\s N} &= \alpha_{\s \d} + \beta_{\s \d} \rho \cos \theta
+ \frac{G}{\rho} \cos \theta.
\end{align}
Here, the constants $\alpha_{\s \u}$, $\alpha_{\s \d}$
are the values of
$\mu_{\s \u}^{\s N}$ and $\mu_{\s \d}^{\s N}$ at $x=0$. Similarly, $\beta_{\s \u}$ and $\beta_{\s \d}$
are $\frac{d\mu^{\s N}_{\s \u}}{dx}$ and $\frac{d\mu^{\s N}_{\s \d}}{dx}$ at $x=0$, which can
be cast in the form
\begin{equation}
\beta_{\s \u (\d)} = \frac{j^0}{2 \sigma^{\s N}}
\left(  1 \pm \mathcal{P} \right).
\end{equation}
With the same accuracy as Eq. (\ref{electric}),
this specification of $\beta_{\s \u}$ and $\beta_{\s \d}$
is valid when $x_0 \gg a$, i.e. when the feedback of the cylinder
on $\mu^{\s N}_{\s \u}$ and $\mu^{\s N}_{\s \d}$ near the
boundary is negligible.

With regard to the net current distribution, the current density is constant
inside the cylinder.  This is, however, not the case for the spin density
distribution, where $\mu_{\s \u}$ and $\mu_{\s \d}$ should be found from
the diffusion equation, Eq. (\ref{diffusion-eq}). It appears that in cylindrical
coordinates we can, similarly to Eq. (\ref{electric}), keep only the solutions
corresponding to the zeroth and
the first angular momenta
\begin{equation}
\mu_{\s \u}^{\s F}
- \mu_{\s \d}^{\s F} = A I_0\left(\frac{\rho}{\lambda_{\s F}}\right)
+ B I_1\left(\frac{\rho}{\lambda_{\s F}}\right) \cos \theta,
\label{eq-deltamu-cyl}
\end{equation}
where $I_0(z)$ and $I_1(z)$ are the modified Bessel functions.
Eq. (\ref{eq-deltamu-cyl}) describes the decay of spin imbalance
upon approaching the center of the cylinder. On the other
hand, it follows from the local current conservation, \hspace{5mm} $\nabla \cdot ({\bm j}_{\s \u} + {\bm j}_{\s \d}) = 0$,
that the combination of $\sigma^{\s F}_{\s \u} \mu^{\s F}_{\s \u} + \sigma^{\s F}_{\s \d} \mu^{\s F}_{\s \d}$
does not decay, i.e.
\begin{equation}
\sigma^{\s F}_{\s \u} \mu^{\s F}_{\s \u} +
\sigma^{\s F}_{\s \d} \mu^{\s F}_{\s \d} =
(\sigma^{\s F}_{\s \u} + \sigma^{\s F}_{\s \d} ) C
+ D \rho \cos\theta.
\end{equation}
The constants $C$, $D$ together with the constants $A$ and $B$ should be found from
the boundary conditions at $\rho = a$.
Matching $\mu^{\s N}_{\s \u}$ and $\mu^{\s F}_{\s \u}$ at $\rho=a$
yields
\begin{align}
C + \frac{\sigma^{\s F}_{\s \d}}{\sigma^{\s F}_{\s \u} + \sigma^{\s F}_{\s \d}} A I_0
\left( \frac{a}{\lambda_{\s F}} \right) &=  \alpha_{\s \u}, \\
\frac{Da}{\sigma^{\s F}_{\s \u} + \sigma^{\s F}_{\s \d}}
+ \frac{\sigma^{\s F}_{\s \d} B}{\sigma^{\s F}_{\s \u} + \sigma^{\s F}_{\s \d}}
I_1\left( \frac{a}{\lambda_{\s F}} \right) &=
\beta_{\s \u} a + \frac{F}{a}.
\label{up-mu}
\end{align}
Similar equations originate from matching $\mu^{\s N}_{\s \d}$ and $\mu^{\s F}_{\s \d}$.

The  condition,  $
\left. \sigma^{\s F}_{\s \u} \frac{\partial \mu^{\s F}_{\s \u}}{\partial \rho} \right|_{\rho =a}
\!=\!\left. \sigma^{\s N} \frac{\partial \mu^{\s N}_{\s \u}}{\partial \rho} \right|_{\rho =a}
$, of the  continuity of the radial current results in
\begin{align}
\label{current-cyl}
\frac{\sigma^{\s F}_{\s \u} \sigma^{\s F}_{\s \d}}{\sigma^{\s F}_{\s \u} + \sigma^{\s F}_{\s \d}}
\frac{A}{\lambda_{\s F}} I_0'\left( \frac{a}{\lambda_{\s F}} \right) &= 0,\\
\frac{\sigma^{\s F}_{\s \u} D}{\sigma^{\s F}_{\s \u} + \sigma^{\s F}_{\s \d}}
+ \frac{\sigma^{\s F}_{\s \u} \sigma^{\s F}_{\s \d}}{\sigma^{\s F}_{\s \u} + \sigma^{\s F}_{\s \d}}
\frac{B}{\lambda_{\s F}} I_1' \left( \frac{a}{\lambda_{\s F}} \right)
&= \sigma^{\s N} \left( \beta_{\s \u} - \frac{F}{a^2} \right).
\label{up-j}
\end{align}
Eqs. (\ref{up-mu}), (\ref{up-j}) express the continuity of the $\cos \theta$ terms in $\mu_{\s \u}$ and ${\bm j}_{\s \u}$. The corresponding conditions for $\mu_{\s \d}$ and ${\bm j}_{\s \d}$ read
\begin{align}
\frac{Da}{\sigma^{\s F}_{\s \u} + \sigma^{\s F}_{\s \d}}
- \frac{\sigma^{\s F}_{\s \u} B}{\sigma^{\s F}_{\s \u} + \sigma^{\s F}_{\s \d}}
I_1\left( \frac{a}{\lambda_{\s F}} \right) &=
\beta_{\s \d} a + \frac{G}{a}, \label{down-mu} \\
\frac{\sigma^{\s F}_{\s \d} D}{\sigma^{\s F}_{\s \u} + \sigma^{\s F}_{\s \d}}
- \frac{\sigma^{\s F}_{\s \u} \sigma^{\s F}_{\s \d}}{\sigma^{\s F}_{\s \u} + \sigma^{\s F}_{\s \d}}
\frac{B}{\lambda_{\s F}} I_1' \left( \frac{a}{\lambda_{\s F}} \right)
&= \sigma^{\s N} \left( \beta_{\s \d} - \frac{G}{a^2} \right). \label{down-j}
\end{align}
Note now, that the four conditions Eqs. (\ref{up-mu}), (\ref{up-j}), (\ref{down-mu}), and (\ref{down-j})
form a closed system of equations for the variables $D, B, F,$ and $G$. Note also, that it is these four variables
which are responsible for the spin current. Solving this system yields
\begin{equation}
\label{spin}
{\bm j}^{\s N}_s({\bm \rho}) = {\bm j}^{\s 0}_s  - \left(a^2\frac{\Pcyl}{\mathcal{P}} \right)\left(  \frac{{\bm j}^{\s 0}_s}{\rho^2} - 2 \frac{({\bm j}^{\s 0}_s \cdot {\bm \rho}) {\bm \rho}}{\rho^4} \right),
\end{equation}
where $\Pcyl$ is defined as
\begin{equation}
\Pcyl  =
\dfrac{\sigma^{\s N} \dfrac{\sigma^{\s F}_{\s \u} - \sigma^{\s F}_{\s \d}}{\sigma^{\s F}_{\s \u} \sigma^{\s F}_{\s \d}}
\gamma_{\s \textit{cyl}}
}
{1 + \dfrac{\sigma^{\s N}}{\sigma^{\s F}_{\s \u} \sigma^{\s F}_{\s \d}} \left(
\dfrac{{\sigma^{\s F}_{\s \u}}^2 + {\sigma^{\s F}_{\s \d}}^2}{\sigma^{\s F}_{\s \u} + \sigma^{\s F}_{\s \d}} \right) \gamma_{\s \textit{cyl}}.
},
\end{equation}
and the constant $\gamma_{\s \mathit{cyl}}$ is given by
\begin{equation}
\gamma_{\s \mathit{cyl}} =
  \left(\dfrac{\lambda_{\s F}}{a} \dfrac{I_1\left[\frac{a}{\lambda_{\s F}} \right]}
  { I_1'\left[\frac{a}{\lambda_{\s F}}\right]} \right).
\label{gamma-cyl-1}
\end{equation}
Equation Eq. (\ref{spin}) represents the spin analog of the charge-current distribution Eq. (\ref{electric}). We see that the spin ``polarizability" of a ferromagnetic cylinder exceeds the electrical polarizability, $a^2$, by a factor $\Pcyl/\mathcal{P}$. For $\sigma^{\s N}\ll \sigma^{\s F}$ this factor simplifies to
\begin{equation}
\label{simplified}
\frac{\Pcyl}{\mathcal{P}}\approx \frac{\gamma_{\s \textit{cyl}}}{\gamma}=\Bigl(\frac{\lambda_{\s N}}{\lambda_{\s F}}\Bigr)\gamma_{\s \textit{cyl}},
\end{equation}
where $\gamma_{\s \textit{cyl}}$, defined by Eq. (\ref{gamma-cyl-1}), depends only on the ratio $a/\lambda_{\s F}$.
Thus, for $\lambda_{\s N} \gg \lambda_{\s F}$, which is the case for
a ferromagnet-semiconductor interface, we find that the spin polarizability of the embedded ferromagnetic  cylinder {\em exceeds substantially the electrical polarizability}.

The above finding can be interpreted as follows. For $\sigma^{\s N}\ll \sigma^{\s F}_{\s \u}, \sigma^{\s F}_{\s \d}$ the expression Eq. (\ref{eq-wyder}) for polarization of the injected current can be viewed as a ratio of two resistances, one having the
resistivity $1/\sigma^{\s N}$ and the length $\lambda_{\s N}$, and the other having the resistivity $\left(1/\sigma^{\s F}_{\s \u}
-1/\sigma^{\s F}_{\s \d}\right)$ and the length $\lambda_{\s F}$.
Then the enhancement of spin polarization predicted by
Eq. (\ref{spin}) can be viewed as a replacement of  $\gamma=\lambda_{\s F}/\lambda_{\s N}$ by the effective ratio $\gamma=\lambda_{\s F}/a$, i.e. the replacement of the length of a ``semiconductor"-resistor by the radius of the cylinder, $a$. This replacement has an origin in inhomogeneity of the electric field on the spatial scale $\sim a$,
similar to the effect of the ``spread resistance".

Generalization of Eqs. (\ref{electric}) and (\ref{spin}) to the case of a ferromagnetic sphere of a radius, $a$, embedded
into a semiconductor is straightforward. The textbook result for the current density distribution reads

\begin{figure}
\includegraphics[width=77mm]{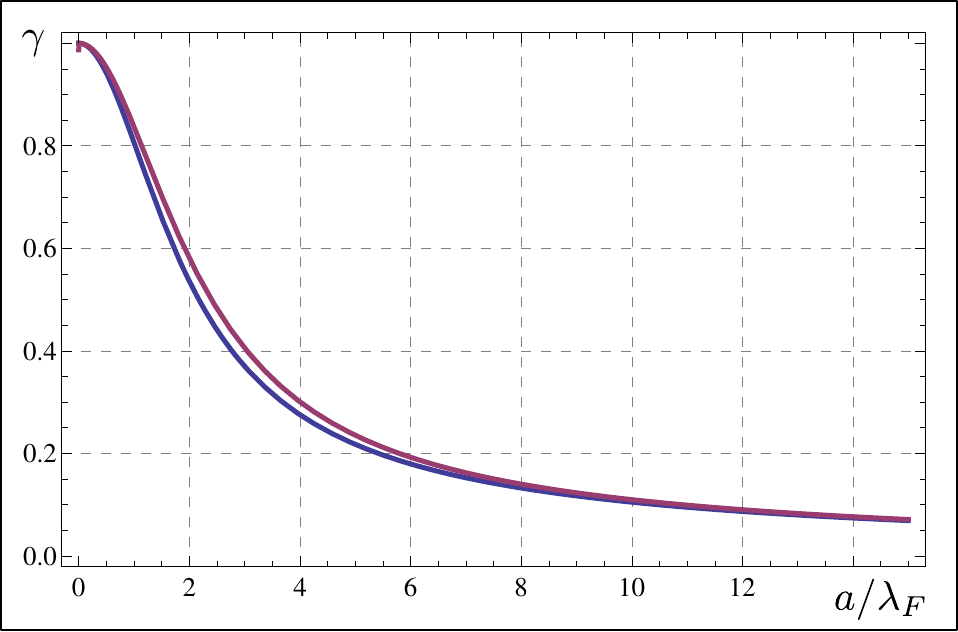}
\caption{(Color online) The spin polarizability of a ferromagnetic cylinder and a sphere exceed the electrical polarizability by
$\frac{\lambda_{\s N}}{\lambda_{\s F}}\gamma_{\s \textit{cyl}}$ and $\frac{\lambda_{\s N}}{\lambda_{\s F}}\gamma_{\s \textit{sphere}}$, respectively. The functions
$\gamma_{\s \textit{cyl}}$ (blue) and $\gamma_{\s \textit{sphere}}$ (purple)
are plotted versus the dimensionless radius, $a/\lambda_{\s F}$, from Eqs. (\ref{gamma-cyl-1}), (\ref{gamma-sphere}).
}
\label{gammas}
\end{figure}

\begin{equation}
\label{electric-sphere}
{\bm j}^{\s N}({\bm r}) = {\bm j}^{\s 0}  - {\bm j}^{\s 0} \frac{a^3}{r^3} + 3a^3 \frac{({\bm j}^{\s 0} \cdot {\bm r}){\bm r}}{r^5},
\end{equation}
while the spin-current density distribution is given by
\begin{equation}
\label{spin-sphere}
{\bm j}^{\s N}_s({\bm \rho}) = {\bm j}^{\s 0}_s
- \left(a^3 \frac{\Psph}{\mathcal{P}} \right) \left(  \frac{{\bm j}^{\s 0}_s}{r^3}
- 3\frac{({\bm j}^{\s 0}_s \cdot {\bm r}) {\bm r}}{r^5}
\right),
\end{equation}
where the induced spin-dipole moment has a form
\begin{equation}
\Psph =
\dfrac{ \frac{3}{2} \sigma^{\s N} \dfrac{\sigma^{\s F}_{\s \u} - \sigma^{\s F}_{\s \d}}{\sigma^{\s F}_{\s \u} \sigma^{\s F}_{\s \d}}
\gamma_{\s \textit{sphere}}
}{1 + 2 \dfrac{\sigma^{\s N}}{\sigma^{\s F}_{\s \u} \sigma^{\s F}_{\s \d}} \left(
\dfrac{{\sigma^{\s F}_{\s \u}}^2 + {\sigma^{\s F}_{\s \d}}^2}{\sigma^{\s F}_{\s \u} + \sigma^{\s F}_{\s \d}} \right) \gamma_{\s \textit{sphere}}.
},
\end{equation}
with $\gamma_{\s \textit{sphere}}$ defined as
\begin{equation}
\gamma_{\s \textit{sphere}} = \left( \frac{\lambda_{\s F}}{a} \frac{i_1 \left[ \dfrac{a}{\lambda_{\s F}} \right]}{
i_1' \left[ \dfrac{a}{\lambda_{\s F}} \right]} \right).
\label{gamma-sphere}
\end{equation}
Here $i_1(z)$ is the modified spherical Bessel function.

Numerically, the functions $\gamma_{\s \textit{cyl}}$  and  $\gamma_{\s \textit{sphere}}$, plotted in Fig. \ref{gammas},
are practically identical. For $a \ll \lambda_{\s F}$, they are equal to $1$, suggesting that the spin
polarization of current at the surface of a small cylinder or a small sphere is given by Eq. (\ref{eq-wyder})
 with the geometrical factor, $\gamma$, equal to $1$ instead of $\lambda_{\s F}/\lambda_{\s N}$.
For  $a \gg \lambda_{\s F}$ both $\gamma_{\s \textit{cyl}}$  and  $\gamma_{\s \textit{sphere}}$ fall off as $\lambda_{\s F}/a$ which correspond to $\gamma\sim \lambda_{\s F}/a$.

\section{Injection from an electrode with a curved interface}

In this Section we will
consider three toy models of spin injection through the interface of finite area.
These models allow exact analytical treatment of a non-planar interface.
They will help us later for the analysis
of spin injection from a ferromagnet into a semiconductor in the presence of interface roughness.

\subsection{``Radial" injection from a cylinder}
\begin{figure}
\includegraphics[width=77mm]{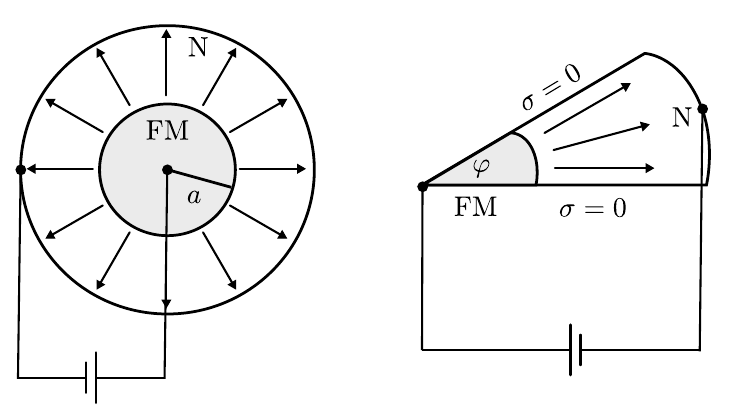}
\caption{Schematic illustration of spin injection from a ferromagnet into a semiconductor in  cylindrical and wedge geometries. The outside radii are much bigger than the radius, $a$, of the ferromagnetic core. }
\label{fig-cylinder}
\end{figure}
By cylindrical geometry we mean the arrangement of ferromagnet and semiconductor
shown in Fig. \ref{fig-cylinder}a.  The cross section of the ferromagnet is a circle
with radius $a$.  Most importantly, the current flows in the plane of the figure rather than along  the axis of the
cylinder, see Fig. \ref{fig-cylinder}a. Obviously, in this geometry, Eq. (\ref{diffusion-eq}) should be solved in the polar coordinates. Namely, if we search for $\mu^{\s F}$ and $\mu^{\s N}$ in the form of the combination of harmonics, $\exp(im\theta)$, then the corresponding radial functions which ensure regular behavior at $\rho\rightarrow 0$ and at $\rho \rightarrow \infty$ are the modified Bessel functions $I_m$ and $K_m$, respectively. Thus we write
\begin{align}
  \mu_{\s \u (\d)}^{\s F} &= \mu_0^{\s F} + \sum_m A_{\s m \u (\d)} I_m(\rho/\lambda_{\s F}) e^{i m \theta}, \\
  \mu_{\s \u (\d)}^{\s N} &= \mu_0^{\s N} + \sum_m B_{\s m \u (\d)} K_m(\rho/\lambda_{\s N}) e^{i m \theta},
\end{align}
Here $\mu_0^{\s F}$ and $\mu_0^{\s N}$ are responsible for the non-polarized part of the current. Similar to the case of an embedded cylinder, the boundary conditions of the continuity of chemical potentials and radial currents should be imposed only on the amplitudes of harmonics. They assume the form
\begin{align}
\label{system1}
\left( A_{\s \u} - A_{\s \d} \right) I_m(a/\lambda_{\s F}) &=
\left( B_{\s \u} - B_{\s \d} \right) K_m(a/\lambda_{\s F}), \\
   \frac{\sigma^{\s F}_{\s \u}}{\lambda_{\s F}} I_m'(a/\lambda_{\s F}) A_{\s \u}
&= \frac{\sigma^{\s N}}{\lambda_{\s N}}         K_m'(a/\lambda_{\s N}) B_{\s \u}, \\
   \frac{\sigma^{\s F}_{\s \d}}{\lambda_{\s F}} I_m'(a/\lambda_{\s F}) A_{\s \d}
&= \frac{\sigma^{\s N}}{\lambda_{\s N}}         K_m'(a/\lambda_{\s N}) B_{\s \d}.
\end{align}
It easily follows from the system Eq. (\ref{system1}) that, for any given $m$, the degree of
polarization has a standard form Eq. (\ref{eq-wyder})  with the factor
 $\gamma$ replaced by $\gamma^{\s r}_{\s \textit{cyl}}$, which is defined as
\begin{equation}
\gamma^{\s r}_{\s \textit{cyl}}(m) =
-\left(\frac{I_m(a/\lambda_{\s F})}{I'_m(a/\lambda_{\s F})}\right)
 \left(\frac{K'_m(a/\lambda_{\s N})}{K_m(a/\lambda_{\s N})}\right) \gamma.
\label{gamma-cyl}
\end{equation}
Consider first the case $m=0$. Then Eq. (\ref{gamma-cyl}) illustrates the
main message formulated in the Introduction.
Namely, when the perimeter, $2 \pi a$, of the F/N boundary exceeds both $\lambda_{\s F}$ and $\lambda_{\s N}$,
then the first ratio in Eq. (\ref{gamma-cyl}) is equal to $1$, while the second ratio becomes $-1$, so that
we get $\gamma^{\s r}_{\s \textit{cyl}} = \gamma$, i.e. the curvature of the boundary has no effect on injection.

Consider now an intermediate domain $\lambda_{\s F} \ll a \ll \lambda_{\s N}$.
Then the argument in the first factor is big,
while the argument in the second factor
is small.  Using the small-$z$ asymptote of $K_0(z)$ we find
\begin{equation}
 \gamma^{\s r}_{\s \textit{cyl}} \Big|_{\lambda_{\s F} \ll a \ll \lambda_{\s N}}
 = \left( \frac{\lambda_{\s N}}{a \ln \frac{\lambda_{\s N}}{a} } \right) \frac{\lambda_{\s F}}{\lambda_{\s N}}= \frac{\lambda_{\s F}}{a\ln \frac{\lambda_{\s N}}{a}}.
 \label{gamma-cyl-mid}
\end{equation}
We conclude that in the intermediate domain the injection efficiency
is enhanced  essentially by $ \lambda_{\s N}/a$.
Note that the above result matches, with logarithmic accuracy, the result Eq. (\ref{gamma-cyl-1}) for the different
geometry in which the external charge and spin currents flow not from, but rather {\em through} the ferromagnetic
cylinder.

For a very small contact area $a \ll \lambda_{\s F}$ the arguments of
the Bessel functions in both factors in Eq. (\ref{gamma-cyl}) are small.
From the asymptote $I_0(z) \approx 1 + \frac{z^2}{4}$ we find
\begin{equation}
 \gamma^{\s r}_{\s \textit{cyl}} \Big|_{a \ll \lambda_{\s F} \ll \lambda_{\s N}}
 = \left( \frac{2  \lambda_{\s N} \lambda_{\s F}}{
 	a^2 \ln \frac{\lambda_{\s N}}{a} } \right) \frac{\lambda_{\s F}}{\lambda_{\s N}}
 \sim \frac{\lambda_{\s F}^2}{a^2}.
 \label{gamma-cyl-tiny}
\end{equation}
In the previous Section we have already realized that  large spin diffusion length, $\lambda_{\s N}$,
disappears from the injection efficiency.
Eqs. (\ref{gamma-cyl-mid}), (\ref{gamma-cyl-tiny}) essentially illustrate the same message and reaffirm the
above picture that for large $\lambda_{\s N}$ the spin resistance of the semiconductor should be replaced by the
spread resistance.

\begin{figure}
\includegraphics[width=77mm]{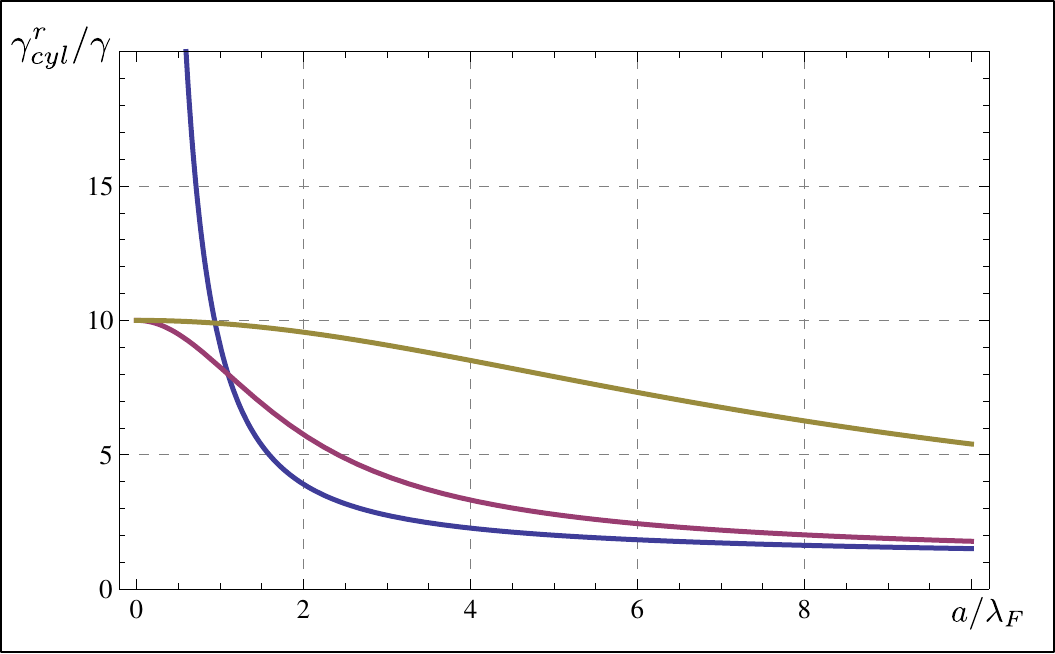}
\caption{(Color online) The enhancement factor of the injected spin polarization
 is plotted for $\lambda_{\s N}=10\lambda_{\s F}$ from Eq. (\ref{gamma-cyl}) versus the dimensionless radius, $a/\lambda_{\s F}$ of the ferromagnetic cylinder for angular momenta $m=0$ (blue), $m=1$ (purple), and $m=6$ (gold). The decay of the enhancement factor corresponds to $\frac{a}{\lambda_{\s F}}\sim m.$ }
\label{fig-gamma}
\end{figure}

For higher azimuthal harmonics, $m$, the enhancement of the
injection efficiency is less pronounced as it is illustrated in
Fig. \ref{fig-gamma}.  On the other hand, for larger $m$, the enhanced
injection efficiency persists over a wider domain of the contact
perimeters, $2 \pi a$.

%
%

\subsection{Injection in the wedge geometry}
Additional insight into the {\em geometrical} enhancement of the spin
injection can be inferred  from the wedge-like arrangement of the F/N
boundary illustrated in Fig. \ref{fig-cylinder}b.  A novel feature
present in Fig. \ref{fig-cylinder}b is that both the ferromagnetic
injector and semiconductor are surrounded by vacuum with $\sigma = 0$.
Then in addition to the conditions Eqs. (\ref{current-condition}),
(\ref{mu-condition}) we must also require that the normal component of the
current at the boundary with the vacuum is zero.  This condition imposes
the following angular dependence of the potentials,
\begin{equation}
\mu_{\s \u} - \mu_{\s \d} \sim \cos \frac{n \pi \theta}{\varphi},
\label{eq-wedge-angle}
\end{equation}
where $\varphi$ is the opening angle of the wedge.
It can be easily checked that the dependence Eq. (\ref{eq-wedge-angle})
translates into the following form of the enhancement factor
\begin{equation}
\gamma_{\s \textit{wedge}} = -\left(\frac{I_{\frac{n \pi}{\varphi}}(a/\lambda_{\s F})}
    {I'_{\frac{n \pi}{\varphi}}(a/\lambda_{\s F})}\right)
	\left(\frac{K'_{\frac{n \pi}{\varphi}}(a/\lambda_{\s N})}
	{K_{\frac{n \pi}{\varphi}}(a/\lambda_{\s N})}\right) \gamma.
\label{gamma-wedge}
\end{equation}
It is easy to see that the small opening angle, $\varphi$,
is completely equivalent to the angular momentum $m=\frac{\pi}{\varphi}$ in
Eq. (\ref{gamma-cyl}).  We have seen above that, for large $m$, the
enhancement is $\frac{\lambda_{\s N}}{\lambda_{\s F}}$ and falls off with
$a$ {\em slowly}, see Fig. \ref{fig-gamma}.  In this sense high values of
$m$ are desirable, but the modes with high $m$ are hard to excite.  The
wedge geometry ``simulates'' high-$m$ values due to the small opening
angle.

\subsection{Injection with ``size quantization"}
 \begin{figure}
\includegraphics[height=75mm]{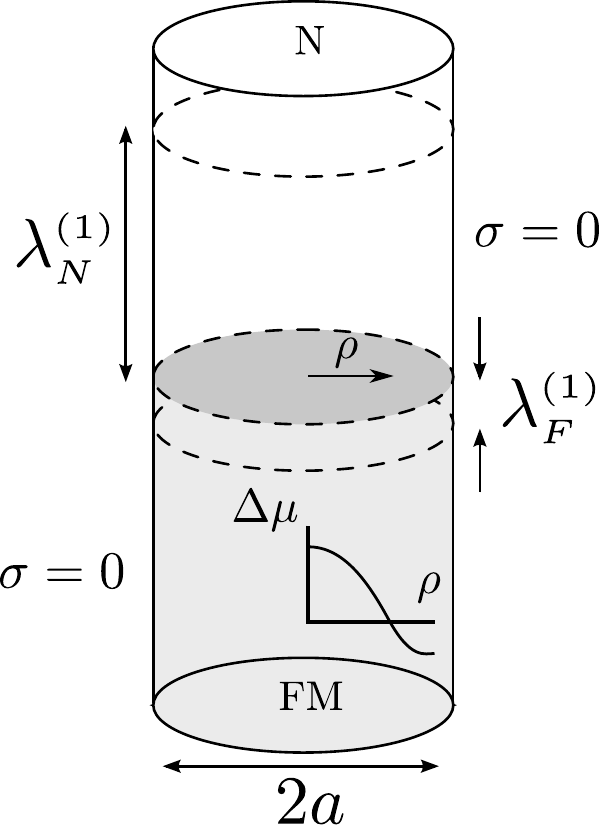}
\caption{ Schematic illustration of the size-quantization effect in spin injection. The modes, corresponding to oscillating distributions
 of $\Delta \mu$  in radial direction, decay along $x$ with decrements larger than $\lambda_{\s F}$ and $\lambda_{\s N}$, see Eqs. (\ref{effective1}), (\ref{effective2}). The behavior of $\Delta \mu (\rho)$ for $n=1$ is illustrated schematically.}
\label{fig-finite}
\end{figure}

In this subsection we study the enhancement of polarized injection for a more
realistic situation when the ferromagnet and semiconducting materials
contact each other over a finite area, $\pi a^2$, as illustrated in
Fig. \ref{fig-finite}a.  We restrict consideration to axisymmetric
solutions, $m=0$.  Then $\Delta \mu$, e.g.  in ferromagnet, satisfies the
equation
\begin{equation}
  \frac{1}{\rho} \frac{\partial }{\partial \rho}
  \left( \rho \frac{\partial}{\partial \rho} \Delta \mu^{\s F} \right) +
  \frac{\partial^2}{\partial z^2} \Delta \mu^{\s F} = \frac{\Delta \mu^{\s F}}
    {\lambda^2_{\s F}}.
\label{diffusion-cyl-z}
\end{equation}
The enhancement in this geometry
emerges from the fact that the condition of absence of current through the sides of the
cylinder imposes ``size quantization'' of the radial dependencies of  the
chemical potentials $\mu^{\s F}$ and $\mu^{\s N}$.  Namely,
$\Delta \mu^{\s F}$ and $\Delta \mu^{\s N}$
behave with $\rho$ as a zero-order Bessel function, $J_0\left( \frac{\alpha \rho}{a}\right)$,
 where  the constant $\alpha$ is
fixed by absence of radial current at $\rho = a$, i.e.
$J_0'(\alpha) = 0$. Thus the values of $\alpha$ are the roots,
$\alpha_{1n}$, of the first-order Bessel function.

What is most important for the enhancement of the injection is that
the
solutions corresponding to different $n$ have different decay decrements
along $z$.  Indeed, from Eq. (\ref{diffusion-cyl-z}) we get
\begin{equation}
\label{effective1}
 \left( \frac{1}{{\lambda^{\s (n)}_{\s F}}}\right)^2 = \frac{1}{\lambda_{\s F}^2}
  + \frac{\alpha_{1n}^2}{a^2},
\end{equation}
for the ferromagnet and similarly
\begin{equation}
\label{effective2}
  \left(\frac{1}{\lambda^{\s (n)}_{\s N}}\right)^2 = \frac{1}{\lambda_{\s N}^2}
  + \frac{\alpha_{1n}^2}{a^2},
\end{equation}
for the semiconductor.  The
modification of $\lambda_{\s F}$ and $\lambda_{\s N}$ modifies the
effective spin-resistances and thus the injection efficiency.
This modification amounts to a replacement of $\lambda_{\s F}$, $\lambda_{\s N}$
in the geometrical factor $\gamma$ in Eq. (\ref{eq-wyder}) by $\lambda^{\s (n)}_{\s F}$ and $\lambda^{\s (n)}_{\s N}$, respectively.
Overall we get
\begin{equation}
  \gamma_{\s n}(a) = \left(
    \sqrt{\dfrac{a^2 + \alpha_{1n}^2 \lambda_{\s N}^2}
	            {a^2 + \alpha_{1n}^2 \lambda_{\s F}^2}}\;
  \right) \gamma.
\label{gamma-size-quantization}
\end{equation}
Again, for $a \gg \lambda_{\s F}, \lambda_{\s N}$ we reproduce the
infinite-area result $\gamma_{\s n}(a) = \gamma$.  For
$\lambda_{\s F} \ll a \ll \lambda_{\s N}$ the result becomes
\begin{equation}
\label{gamma(a)}
  \gamma_{\s n}(a) = \left(\frac{\alpha_{1n} \lambda_{\s N}}{a} \right)
  \gamma \sim \frac{\lambda_{\s F}}{a}.
\end{equation} Similar to Eq.  (\ref{gamma-cyl-mid}) the largest
spin-diffusion length $\lambda_{\s N}$ drops out of the spin-injection
efficiency.  In other words, the spin-resistance of the semiconductor is
determined by a cylinder of diameter $a$ and height $\sim a$.

It is seen from Eq. (\ref{gamma(a)}) that the bigger is $n$ the stronger
is the enhancement. On the other hand, similar to the solutions with high angular momentum,
the solutions with large radial number are hard to excite. In the absence of inhomogeneity
in the $\rho$-direction, the dominant solution is $\Delta\mu =\text{const}(\rho)$, and hence  no enhancement
of the injection.


\section{Realistic geometries}

\subsection{Rough interface: stalactites}
\begin{figure}
\includegraphics[width=77mm]{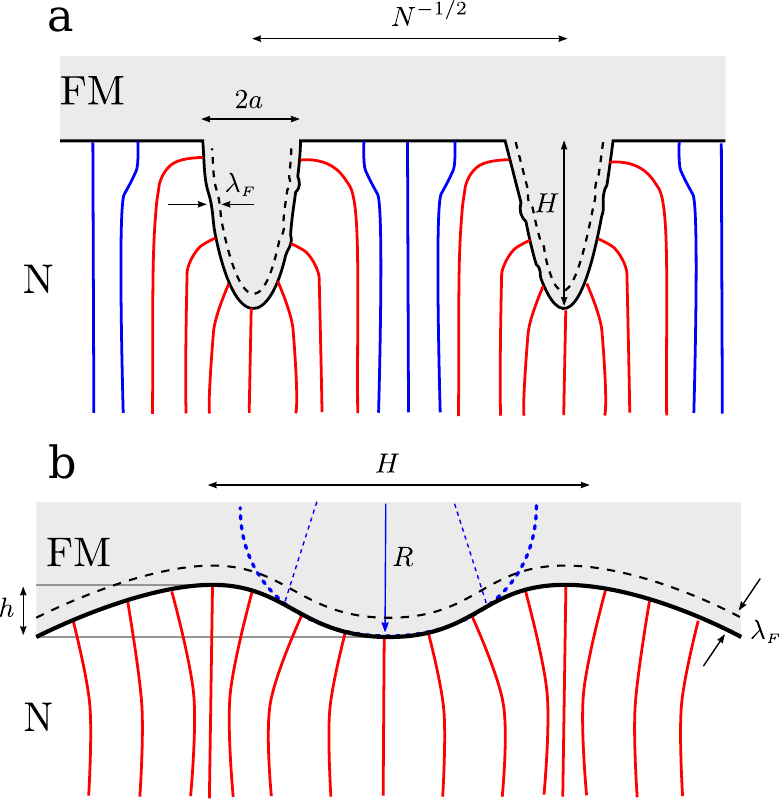}
\caption{(Color online) Two models of a rough interface are depicted schematically; (a)interface roughness is modeled as a system of ``stalactites" with surface density, $N$, and linear sizes $H$ and $a$. For $H\gg a$ the current is injected radially from the stalactite surface, so that the enhancement factor is given by Eq. (\ref{gamma-cyl-mid}). At distance $\sim H$ from the top of the stalactite the current distribution is homogeneous. (b) a model of an
interface roughness with characteristic period, $H$, and amplitude
$h \ll H$. Due to smallness of the ratio $h/H$ the characteristic radius
of curvature $R=H^2/h$ is much bigger than $H$.}
\label{rough}
\end{figure}
We will consider two models of the F/N interface roughness, see Fig. \ref{rough}. In the first model the ferromagnet
penetrates in a stalactite fashion into the semiconductor. We will assume
that the height, $H$, and the diameter, $a$,  of a typical stalactite  is much bigger than $\lambda_{\s F}$ but much smaller than $\lambda_{\s N}$. We start from the simplest case $H=a$. It follows from the above consideration that even when the surface density, $N$, of stalactites is small, $Na^2\ll 1$ their presence might cause a strong net enhancement of the spin injection.  As it was explained above, the origin of the enhancement is that for $\sigma_{\s \u}^{\s F}, \sigma_{\s \d}^{\s F} \gg \sigma^{\s N}$ the force lines of electric field are normal to the curved F/N interface and the magnitude of the induced field decays at a scale $\sim a$ away from the interface. For the purposes of the spin injection it is also important that at distances from the stalactite surface also $\sim a$ the field lines become straight. This
ensures that the contributions to the net spin polarization from the plane region of the interface and from the stalactites are additive. The last crucial observation is that, as long as the distance from the interface remains smaller than $\lambda_{\s N}$, the polarization of the current injected from the surface of the stalactite remains the same. Summarizing all the above arguments, we can present the average polarization as
\begin{align}
\label{average}
\mathcal{P}_{\textit{injected}} &= \mathcal{P} \left( 1 + N \frac{\mathcal{P}_{\s \textit{stal}}}{\mathcal{P}}a^2 \right) \\
&\approx \sigma^{\s N} \left(\frac{1}{\sigma^{\s F}_{\s \d}} - \frac{1}{\sigma^{\s F}_{\s \u}} \right)
\frac{\lambda_{\s F}}{\lambda_{\s N}}
\left( 1 + Na^2 \frac{\gamma_{\s \textit{stal}}}{\gamma} \right),
\end{align}
where $\mathcal{P}_{\s \textit{stal}}$ and  $\gamma_{\s \textit{stal}}$ in the second term in the brackets stand for the
spin polarizability and the enhancement factor of the  stalactite, respectively. In the case of a sphere and a cylinder, and all the cases considered thereafter, we obtained that, with accuracy of a numerical factor, the result for the enhancement factor is the same. For this reason we set $\gamma_{\s \textit{stal}}/{\gamma} \sim \lambda_{\s N}/a$. This leads us to the final result
\begin{equation}
\label{final}
\mathcal{P}_{\textit{injected}} =  \sigma^{\s N} \left(
\frac{1}{\sigma^{\s F}_{\s \d}} - \frac{1}{\sigma^{\s F}_{\s \u}} \right)\frac{\lambda_{\s F}}{\lambda_{\s N}}
\left( 1 + Na \lambda_{\s N} \right).
\end{equation}
It is seen from Eq. (\ref{final}) that the stalactites dominate the injection when the distance between the neighbors
is smaller than  $(\lambda_{\s N}a)^{1/2}$. This condition is compatible with the assumption that this distance is bigger than $a$.

We now turn to the limit $H\gg a$. In this limit, the condition, $\sigma^{\s F}\gg \sigma^{\s N}$ leads to the field enhancement near the stalactite interface, in order to turn the field lines to from vertical to horizontal, see Fig. \ref{rough}. As a result the polarized current is injected in the radial direction. Thus for $\lambda_{\s F}\ll a$ we can use Eq. (\ref{gamma-cyl-mid}) for the
enhancement factor. Since the current from the stalactites is injected radially, while from the rest of the interface it is injected normally, it is convenient to calculate the   average  polarization
using the cross section a distance $\sim H$ below the stalactites. Indeed, at a distance $\sim H$ below the stalactites  the net current becomes homogeneous. A nontrivial element of the calculation is that,  at such distances, the current lines, emanating from a given stalactite, occupy the area $\sim H^2$. In other words, the current
injected from the area $2\pi aH$, which is the surface area of the
stalactite, spreads out into the area $\sim H^2$. This is indeed the
case, since, due to enhancement near the interface, the radial electric field exceeds the field away from the interface by $\sim H/a$. Basing on the above remarks, we conclude that the generalization of Eq. (\ref{average}) to the case $H\gg a$ reads
\begin{align}
\mathcal{P}_{\textit{injected}} &= \mathcal{P} \left( 1 + N \frac{\mathcal{P}_{\s \textit{stal}}}{\mathcal{P}}H^2 \right) \\
&\approx \sigma^{\s N} \left(\frac{1}{\sigma^{\s F}_{\s \d}} - \frac{1}{\sigma^{\s F}_{\s \u}} \right)
\frac{\lambda_{\s F}}{\lambda_{\s N}}
\left( 1 + NH^2 \frac{\gamma_{\s \textit{stal}}}{\gamma} \right).
\end{align}
Substituting $\gamma_{\textit{stal}}$ from
Eq. (\ref{gamma-cyl-mid}), we arrive at the final result
\begin{equation}
\mathcal{P}_{\textit{injected}} =  \sigma^{\s N} \left(
\frac{1}{\sigma^{\s F}_{\s \d}} - \frac{1}{\sigma^{\s F}_{\s \u}} \right)\frac{\lambda_{\s F}}{\lambda_{\s N}}
\left( 1 + N \frac{H^2 \lambda_{\s N}}{a} \right).
\label{final1}
\end{equation}
Note that, while the condition $Na^2\ll 1$ ensures that the neighboring stalactites do not overlap, Eq. (\ref{final1}) is only valid for $N \lesssim 1/H^2$. The physical reason for this is that for
higher densities the field enhancement takes place only within the distance $\sim N^{-1/2}$ from the tips of stalactites. This is because the stalactites screen the external field {\em collectively}\cite{tubes}. Thus for $N \gtrsim 1/H^2$ the result Eq. (\ref{final1}) saturates.

\subsection{Rough interface: small roughness amplitude}

 In the second model of a rough F/N interface the interface profile is sinusoidal, see Fig. \ref{rough} with  characteristic amplitude, $h$, and the period, $H$.  To find the  $\gamma$-parameter, $\gamma_{\s \textit{sin}}$, for this
 model, we notice that the effective radius of curvature corresponding to a given element of the interface is $R=H^2/h$. As illustrated in Fig. \ref{rough}b, one can view the injection from this element as a radial injection from the surface of a cylinder with radius $R$. This suggests that we can estimate $\gamma_{\s \textit{sin}}$ with the help of  Eq.  (\ref{gamma-cyl-mid}) in which the radius $a$ is replaced by $R$. This yields
 \begin{equation}
\label{gamma-sin}
\gamma_{\s \textit{sin}} \sim \frac{\lambda_{\s F} h}{H^2}.
\end{equation}
The corresponding expression for polarization of the injected current reads

\begin{equation}
\label{polarization-sin}
\mathcal{P}_{\s \textit{sin}} = \sigma^{\s N}
\left( \frac{1}{\sigma^{\s F}_{\s \d}} - \frac{1}{\sigma^{\s F}_{\s \u}}\right)\frac{ \lambda_{\s F} h}{H^2}.
\end{equation}

Note that the result Eq. (\ref{gamma-cyl-mid}) pertained to the intermediate
domain of the radii, $a$, namely $\lambda_{\s F} \ll a \ll \lambda_{\s N}$.
With $a$ replaced by $R$, this domain turns out to be
 $(  \lambda_{\s F}h)^{1/2} \ll H \ll ( \lambda_{\s N}h)^{1/2}$. The first inequality suggests that $\gamma_{\s \textit{sin}}$ is small. As the period, $H$,
 gradually decreases and reaches $(\lambda_{\s F}a)^{1/2}$,
 the value $\gamma_{\s \textit{sin}}$ reaches $1$ and saturates upon further decrease of $H$.

\subsection{Injection via a ferromagnetic pillar}

\begin{figure}
\includegraphics[width=77mm]{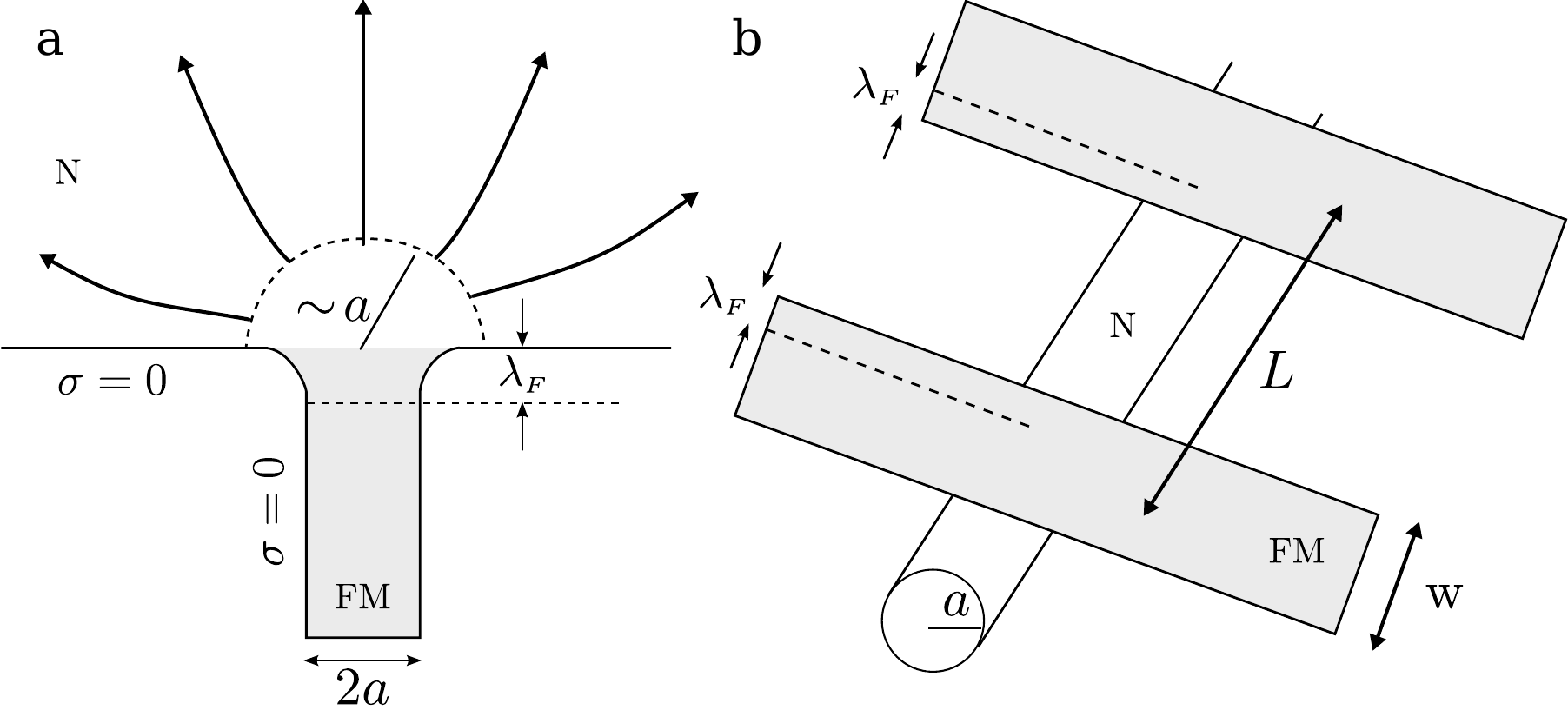}
\caption{(a) Schematic illustration of spin injection through a nanopillar with radius $a\ll \lambda_{\s F}$. The force
lines of electric field a curved within a distance $\sim a$ from the interface; (b) A typical nonlocal geometry employed for measurement of spin injection. Ferromagnetic injector and detector of width, w, are separated by distance, $L$,
on top of a semiconduction wire with radius, $a$. It is presumed that the sizes are ordered as follows:
$\lambda_{\s F} \ll a \ll \text{w} \ll L \ll \lambda_{\s N}$.  }
\label{needle}
\end{figure}

Assume that a ferromagnetic pillar of a radius, $a$, is in contact with a semiconductor surface, as shown in Fig. \ref{needle}a. Since the conductivity
in the lower half-space, except for the interior of the  pillar is zero, the
current lines near the interface between a semiconductor and  non-conducting medium are almost horizontal. In general, the electric  field
inside the semiconductor  behaves as a field of a charge distributed over the F/N interface. At distances $\gtrsim a$ from the contact this charge can be replaced by a point charge, so that electric field falls off with distance $r$
from the pillar as $1/r^2$. This knowledge is sufficient to deduce the electrical spread resistance of the contact to be $1/2 \pi \sigma^{\s N}a$. Naturally, the spin resistance is the same. The spin resistance of the ferromagnet is calculated taking into account that the cross section  area
is $\pi a^2$, and the length is $\lambda_{\s F}$, yielding the value
$\left(\frac{1}{\sigma^{\s F}_{\s \d}} - \frac{1}{\sigma^{\s F}_{\s \u}}\right) \frac{\lambda_{\s F}}{\pi a^2}.$
Then the spin polarization of the injected current calculated as a ratio of
the two spin resistances reads
\begin{equation}
\mathcal{P}_{\s \textit{pillar}} =  \sigma^{\s N}
\left(\frac{1}{\sigma^{\s F}_{\s \d}} - \frac{1}{\sigma^{\s F}_{\s \u}}\right) \frac{\lambda_{\s F}}{2 a}.
\label{polarization-pillar}
\end{equation}
Note that this result is in line with the above expressions for $\mathcal{P}$, e.g.  Eq.  (\ref{gamma-cyl-mid}), if we treat $a$ as a radius of curvature.
Like in previous examples, Eq. (\ref{polarization-pillar}) applies when
the ratio $\lambda_{\s F}/ a$ is small.  As $\lambda_{\s F}$ exceeds $a$, the ratio should be replaced by $1$.

\section{discussion}
\begin{itemize}

\item
The main outcome of our study is that for a curved F/N interface the
ratio $\lambda_{\s F}/\lambda_{\s N}$ in the seminal expression, Eq. (\ref{eq-wyder}), for the polarization of the injected current should be replaced by $\lambda_{\s F}/R$, where $R$ is the local curvature of the interface. Naturally, this replacement is appropriate when $R$ is smaller than $\lambda_{\s N}$.

In our study we focused on a single F/N boundary. Note
that,
shortly after the publication of the paper Ref. \onlinecite{wyder}, it was proposed\cite{SilsbeeTrueGeometry} to measure the spin injection in a non-local geometry where the ferromagnetic injector and detector are spatially separated in the lateral direction. Nowadays this geometry is absolutely common in experimental studies of the spin
transport, see e.g.  recent papers Refs. \onlinecite{InjectionIntoNanowire,vanWeesGrapheneInjection,LatestFertNonlocal}.
At the same time, the theories, see e.g. Refs.
\onlinecite{CrossedGeometry1DCalculation},
\onlinecite{Flatte-Yu, Hershfield, JapaneseInjection},
used to describe these non-local spin-injection setups
are, essentially, ``one-dimensional.''
More specifically, they assume that the injection takes place along $x$ while the subsequent propagation
of spin-polarization happens along $y$, so that these two processes are decoupled.

To emphasize this point, in Fig. \ref{needle}b. the nonlocal geometry
is illustrated schematically, and all relevant sizes are indicated.
Theories \onlinecite{CrossedGeometry1DCalculation},
\onlinecite{Flatte-Yu, Hershfield, JapaneseInjection} treat the injection
in this geometry by dividing it into two F/N junctions in sequence. In particular the spin resistance of the wire part in  Fig. \ref{needle}b.
is set to be proportional\cite{CrossedGeometry1DCalculation} to the wire length, $L$.

We argue that such approach completely neglects the curving of the current paths upon injection from the ferromagnet into the wire, and that this curving changes dramatically the injected polarization. Our answer for this polarization contains $\lambda_{\s F}/\text{w}$ and does not depend on $L$.
This is because the field lines turn 90 degrees over the distance $\sim \text{w}$, so that $\text{w}$ plays the role of the radius of curvature.

In fact, the importance of curving of the current paths in nonlocal geometry for for calculation of nonlocal electrical resistance  was pointed out in 
Refs. \onlinecite{JapaneseCurrentDistribution}, \onlinecite{SilsbeeWire}.

\item
The idea that decreasing the contact area between a ferromagnet and a semiconductor can enhance the efficiency of the
spin injection was previously expressed in Refs. \onlinecite{GeomtericNanopillarTheory}, \onlinecite{GeometricalEffectsTheory}. In particular, the geometry considered in Ref. \onlinecite{GeomtericNanopillarTheory} was a ferromagnetic nanopillar
with a radius $2\text{ nm}$. Numerical simulations\cite{GeomtericNanopillarTheory} suggest that constricting the injector to a small-radius cylinder  leads to the enhancement of polarization from $10^{-3}\%$  to $\sim 1\%$ for certain parameters of a ferromagnet and semiconductor. Surprisingly, in interpreting the simulation results the authors of Ref. \onlinecite{GeomtericNanopillarTheory} contend that the spread resistance suppresses the injection. In Ref. \onlinecite{GeometricalEffectsTheory} the numerical simulations were also performed for a nanopillar-injector geometry.
In addition to Ref. \onlinecite{GeomtericNanopillarTheory} the authors traced a gradual enhancement of injection with
decreasing the nanopillar area. However, another numerical finding reported in Ref. \onlinecite {GeometricalEffectsTheory}, that the efficiency increases with increasing of the semiconductor thickness, seems counterintuitive.

\item
Experimental results of Ref. \onlinecite{dali} seem to offer a partial support of our predictions. In Ref. \onlinecite{dali} it was demonstrated that the performance of the vertical organic spin  valves
 consisting of an organic layer of thickness $D \sim 100 \text{ nm} $ sandwiched between two ferromagnets,
  $\text{Co}/\text{Alq}_3/\text{LSMO}$, can be significantly improved by  covering the Co electrodes by a closely packed layer of Co nanodots.
These nanodots represented spheres of a radius $\approx 1 \text{ nm}$ which is  smaller than $\lambda_{\s F}\approx 59 \text{ nm}$ for Co, the value accepted in the literature\cite{InjectionIntoNanowire}.
Incorporation of nanodots allowed the authors to increase the magnetoresistance from $\sim 10\%$ to $\sim 300\%$.
 Their explanation was that the layer of nanodots eliminates the ``ill-defined" organic spacer layer\cite{valy1,valy2}. According to arguments presented above, the increase of injection efficiency due to the
 curvature of surface of nanodots is $\sim D/\lambda_{\s F}$.

%
%
%

\end{itemize}

\section{Acknowledgements}
  We are grateful to Z. V. Vardeny
for piquing our interest in the subject. This work
was supported by NSF through MRSEC DMR-1121252.

\end{document}